\documentclass[aps,pra,twocolumn,showpacs,floatfix]{revtex4}

\usepackage{graphicx}% Include figure files
\usepackage{dcolumn}% Align table columns on decimal point

\usepackage{bm}
\usepackage{epsfig}

\usepackage{epsf}

\begin{document}

\title{Vortex lattices in Bose-Einstein condensates with Dipolar Interactions Beyond the Weak Interaction Limit}
\author{S. Komineas$^{1,2}$ and  N. R. Cooper$^1$}
\affiliation{$^1$Theory of Condensed Matter Group, Cavendish Laboratory,
J.J. Thomson Avenue, Cambridge CB3 0HE, United Kingdom  \\
$^2$ Max-Planck Institute for the Physics of
Complex Systems, N\"othnitzer Str. 38, 01187, Dresden, Germany.}

\date{\today}

\begin{abstract}
We study the ground states of rotating atomic Bose-Einstein
condensates with dipolar interactions.  We present the results of
numerical studies on a periodic geometry which show vortex lattice
ground states of various symmetries: triangular and square vortex
lattices, ``stripe crystal'' and ``bubble crystal''.  We present the
phase diagram (for systems with a large number of vortices) as a
function of the ratio of dipolar to contact interactions
and of the chemical potential.
We discuss the experimental requirements for observing transitions
between vortex lattice groundstates via dipolar interactions.  We
finally investigate the stability of mean-field supersolid phases of a
quasi-2D {\it non}-rotating Bose gas with dipolar interactions.
\end{abstract}

%03.75.Kk Dynamic properties of condensates; collective and hydrodynamic
%excitations, superfluid flow
%05.30.Jp Boson systems (for static and dynamic properties of
%Bose-Einstein condensates, see 03.75.Hh and 03.75.Kk)

%\pacs{03.75.Lm, 03.75.Kk, 73.43.Cd}
\maketitle

\section{Introduction}

One of the most dramatic manifestations of the collective quantum
behaviour of a Bose-Einstein condensate is its unusual response to
rotation.  The formation of quantised vortex lines, around which the
phase of the condensate wavefunction changes by $2\pi$, is a direct
consequence of the existence of a collective phase-coherent quantum
state. The further ordering of quantised vortices at low temperatures
into regular vortex lattices is a signature not just of Bose-Einstein
condensation but of the presence of non-zero (repulsive) interactions
between the constituent particles.

Techniques for rotating condensates in ultra-cold atomic gases are now
well established, and have led to very beautiful demonstrations of the
physics of quantised vortices in these systems.
Experimental studies of rapidly rotating atomic Bose gases have shown
evidence for the formation of ordered arrays of very large numbers of
quantised vortices,\cite{bretin04,aboshaeer01,schweikhard04} and detailed
studies of the structure and dynamics of these vortex lattices have
been performed.
These studies are consistent with the expected properties for a Bose
gas interacting with short range contact interactions (representing
the van der Waals' forces), for which the rotating
groundstate is a triangular vortex lattice.

The achievement of Bose-Einstein condensation of
chromium-${52}$\cite{griesmaier05} opens up the possibility of experimental
studies of condensates with long-range interactions.  Chromium has a
large permanent magnetic dipole moment, which leads to significant
dipolar interactions in addition to the usual short-range
interactions.
It is of interest to ask how these long-range interactions affect the
vortex lattice groundstates of the rotating condensate.
It has been shown\cite{nigel05} that the non-local interactions
arising from dipolar interactions can lead to changes in the structure
of the vortex lattices. The triangular vortex lattice can be replaced
by vortex lattices of different symmetries: square lattice, ``stripe
crystal'' and ``bubble crystal'' phases.
(For a related study of square and stripe crystals see Ref.\onlinecite{zhang05}.)
The vortex lattice structure is therefore a sensitive indication of
the form of the interparticle interaction.
However, the results of Ref.~\onlinecite{nigel05} are restricted to the
limit of weak interactions, when the mean interaction energy is
smaller than the level spacing of the harmonic trap and the single
particle states are restricted to the lowest Landau level
(LLL)\cite{butts99,wilkin98}.

In this paper we extend the results of Ref.~\onlinecite{nigel05} 
beyond the weak interaction regime, and determine the vortex lattice
structures to be found for arbitrarily strong interactions.  For
sufficiently weak interactions we recover the results of
Ref.~\onlinecite{nigel05};
we find groundstates that are the square
lattice, ``stripe crystal'' and ``bubble crystal'' phases predicted in
Ref.~\onlinecite{nigel05} in the LLL. We determine the effects of
Landau level mixing on these states. We show that for strong
interactions, these new vortex lattice states are unstable to collapse
of the condensate.
%This is consistent with the expected behavior in the limit of strong
%interactions where there are logarithmic repulsions between vortices
%and thus the triangular lattice is stabilized.
Our results establish the conditions required in
order to observe transitions in the vortex lattice groundstate driven
by dipolar interactions.

We also describe the connections with possible ``supersolid'' states
of the {\it non}-rotating gas, and discuss the stability of these states.

\section{Formulation}

We consider a BEC of atoms with mass $M$ which are confined in a
harmonic trap with cylindrical symmetry about the $z$ axis.  We denote
the trap frequencies by $\omega_\|$ and $\omega_\perp$ in the axial
and transverse directions, and the associated trap lengths by
$a_{\|\perp}\equiv \sqrt{\hbar/(M\omega_{\|\perp})}$.  The atoms are
taken to interact through both contact interactions and dipolar
interactions, with a potential
\begin{equation}  \label{eq:potential}
V(\bm{r}) = g\,\delta^3(\bm{r})
 + C_{\rm d}\, V_{\rm d}(\bm{r}),
\end{equation}
where $g$ parameterizes the contact interactions, and $C_{\rm d}$ the
strength of the dipole interaction potential, $V_{\rm d}(\bm{r})$.  We
consider the atomic dipole moments to be aligned with the $z$ axis, as
in the recent experiments \cite{griesmaier05}, which sets the
functional form for $V_{\rm d}({\bf r})$.  Since the classical
dipole-dipole interaction can include also a delta-function
contribution\cite{jackson}, for clarity we define the dipolar
interaction potential by 
%\begin{eqnarray*}
%V_{\rm d}({\bm r}\neq 0) & \equiv & 
%\frac{\partial^2}{\partial z^2} \frac{1}{\sqrt{x^2+y^2+z^2}}\\
% & = & \frac{x^2+y^2-2z^2 }{(x^2+y^2+z^2)^{5/2}}
%\label{eq:vd}
%\end{eqnarray*}
\begin{equation}
V_{\rm d}({\bm r}\neq 0) \equiv
 \frac{x^2+y^2-2z^2 }{(x^2+y^2+z^2)^{5/2}}
\label{eq:vd}
\end{equation}
in which we {\it exclude} the point ${\bm r}=0$, such that any contact
contribution of the dipole interaction is contained in the term proportional
to $g$ in Eq.~(\ref{eq:potential}).
We consider the ratio $g/C_{\rm d}$ to be a parameter that can be varied.
For example in the chromium condensate $g$ can be varied
experimentally by using a
Feshbach resonance \cite{griesmaier05,werner05}.
The system can remain stable even for a small attractive
contact interaction ($g <0$). (We note, however, that the $q=0$
Fourier transform of the effective potential, $\tilde{V}_{q=0}$, should remain
positive.)

The results of Ref.~\onlinecite{nigel05} suggest that the mean field
approximation is excellent even for very dilute systems which may
result from fast rotating gases.  We thus employ here the
Gross-Pitaevskii model.  
We consider a system, rotating with angular frequency $\Omega$, and
containing a large number of vortices, such that near the center of
the trap the particle density varies weakly over the scale of the
vortex spacing.
In the rotating frame, the Hamiltonian becomes equivalent to
the Hamiltonian of a particle of charge $q$ in a uniform magnetic
field $B$, with $qB/M = 2\Omega$\cite{nigel01,ho01}.  We discretise
space and write the kinetic energy for this system in the form \cite{nagaosa}
\begin{equation}  \label{eq:kinetic}
E_{\rm kin} = -  t \sum_{\langle i,j\rangle }\Psi_i^*\Psi_j\,e^{i\phi_{ij}} + {\rm c.c.},
\end{equation}
where $\langle i,j\rangle $ denotes the nearest neighbours of a square
lattice with lattice constant $a$.
%(every pair ($i,j$) is counted only once in the sum).
In our calculations, the lattice
constant $a$ is chosen sufficiently small as to represent the
continuum limit, with the particle mass $M = t a^2/\hbar^2$.
The ``Peierls'' phases are defined by the integral
\begin{equation}  \label{eq:peierls}
\phi_{ij} = \frac{q}{\hbar} \int_{\bm{r}_i}^{\bm{r}_j} \bm{A}\cdot d\bm{l}
\end{equation}
calculated between the positions of the lattice sites $i$ and $j$.  We
choose the Landau gauge for the vector potential $\bm{A} =
B\,x\,\bm{\hat{y}}$, which is convenient in our numerical
calculations.  

We determine the properties of a large array of vortices by studying
the system in a periodic geometry.
Imposing periodic boundary conditions under magnetic translations
requires that
\begin{equation}  \label{eq:nvortices}
L_x L_y = 2\pi\,\ell^2\,N_{\rm v},
\end{equation}
where $N_{\rm v}$ is the number of vortices in the simulation cell
with dimensions $L_x, L_y$.  The magnetic length $\ell=\sqrt{\hbar/qB}
= \sqrt{\hbar/2M\Omega}$ arises as the natural length scale in this
system.  Since we consider a system containing a large number of
vortices, we have $\Omega \simeq \omega_\perp$\cite{footnote1}
so $\ell \simeq a_\perp/\sqrt{2}$.
Vortex lattices in periodic systems have been studied previously
for the case of contact interactions \cite{cozzini06}.

Throughout this paper,
we work in a quasi-two-dimensional (2D) regime, valid when the
confinement in the longitudinal direction is sufficiently strong that
the mean interaction energy (which is on the order of the chemical
potential $\mu$\cite{footnote2}
is small compared to
the level spacing $\hbar\omega_\parallel$. We shall, however, allow
mixing of Landau levels for the in-plane motion, so shall work in the
regime $\hbar \omega_\parallel \gg \mu \sim \hbar\omega_\perp$ which
implies $a_\parallel^2/a_\perp^2\ll 1$, {\it i.e.} that the trap is
{\it oblate}.
In the quasi-2D limit, $\mu \ll \hbar\omega_\parallel$, the condensate
wavefunction is a gaussian along the rotation axis, {\it i.e.},
$\Psi(x,y,z) = \psi(x,y)\, e^{-z^2/2a_\|^2}/(\pi^{1/4}a_\|^{1/2})$.
Integrating the dipolar potential energy along the $z$-axis one
obtains the effective dipolar interaction in two dimensions
\begin{widetext}
\begin{eqnarray}  \label{eq:V2D}
V^{\rm 2D}_{\rm d} (\rho) & \equiv & \int\int V_{\rm d}(\rho, z_1-z_2)\,
     \frac{e^{-(z_1^2+z_2^2)/a_\|^2}}{\pi\,a_\|^2}\; dz_1 dz_2 
   =  \frac{1}{2\sqrt{2\pi}\, a_\|^3}\,
  e^{\zeta^2/4}\,
[(2+\zeta^2)\, K_0(\zeta^2/4) - \zeta^2 K_1(\zeta^2/4)\,],
\end{eqnarray}
\end{widetext}
%\begin{equation}  \label{eq:V2D}
%V^{\rm 2D}_{\rm d} (\rho) 
%   =  \frac{1}{2\sqrt{2\pi}\, a_\|^3}\,
%  e^{\zeta^2/4}\,
%[(2+\zeta^2)\, K_0(\zeta^2/4) - \zeta^2 K_1(\zeta^2/4)\,],
%\end{equation}
where $\rho\equiv\sqrt{x^2+y^2}$, $\zeta\equiv \rho/a_\|$, and $K_n$
are modified Bessel functions.  When $\rho$ is large compared to
$a_\|$, the expression (\ref{eq:V2D}) tends to $1/\rho^3$, which is
the dipolar energy for a 2D monolayer.  We calculate
$1/\rho^3$ by ``Ewald summation'' for the present case of a {\it
periodic} configuration of dipoles \cite{weiss03}, and the remainder
$V^{\rm 2D}_{\rm d} (\rho) - 1/\rho^3$ by direct summation
within an area with radius several times the length $a_\|$ (this
combination is rapidly decaying with $\rho$ so it converges quickly).

For the energy minimization we
use a variant of a norm-preserving
relaxation algorithm which capitalizes on a virial
relation \cite{papanicolaou05} in order to fix the wavefunction norm. 
This is fed with an initial guess and is iterated until
an energy minimum is reached.
In order to investigate the vortex lattice groundstates, we initially
simulate relatively large systems containing $N_{\rm v} =8$ or $16$
vortices. (Typical numerical grid sizes are $N_x, N_y \sim 50 - 60$.)
From these calculations, we find the range of values of the parameters
over which the system is stable against collapse. We further identify
the symmetries of the vortex lattice groundstates that appear,
starting from different initial conditions to give unbiased searches
of vortex lattices of different symmetry-types. 
We subsequently proceed to a systematic calculation of
the energy for each of the above mentioned vortex lattice phases
by using an appropriate simulation cell that is commensurate with the 
translational symmetry of that phase. We use a square cell
in order to obtain a square vortex lattice
(e.g., $N_x=10, N_y=10, N_{\rm v}=1$),
a cell with $N_x/N_y=\sqrt{3}$ for a triangular vortex lattice
(e.g., $N_x=11, N_y=19, N_{\rm v}=2$), and accordingly for the
other vortex lattice phases.
The typical lattice spacing,
implied by Eq.~(\ref{eq:nvortices}),
is $\Delta x = \Delta y = 0.25 \ell$.
We find the energy of each type of vortex lattice 
as a function of the parameter $g/C_{\rm d}$, for various interaction
strength values measured by the chemical potential $\mu$. 
\cite{footnote3}

\section{Vortex lattices}

\begin{figure}
\epsfig{file=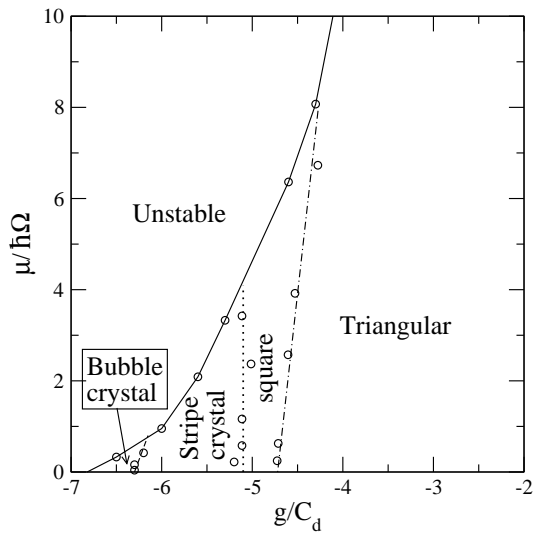,width=7cm}
  \caption{The phase diagram for a trap asymmetry
$a_\|/\ell=0.4$.
The horizontal axis gives the ratio of the contact to dipolar 
interactions and the vertical axis is
the chemical potential (normalized to the energy corresponding
to the rotation frequency $\Omega$ of the condensate).
The condensate is unstable to collapse on the left of the solid line.
Marked on the figure are
the regions where vortex lattices of different symmetries
are the ground state of the system.
%The condensate is unstable in the region left of the solid line.
%The triangular vortex lattice is stable on the right of the dashed line.
%Between the dashed and the dashed-dotted line we have a square
%vortex lattice.
%Between the dashed-dotted line and the dotted line we have a
%``stripe crystal'' vortex phase. To the left of the dotted
%line we find a ``bubble crystal'' phase.
The lines are fits to numerical results marked by circles.
}
\label{fig:phasediagram}
\end{figure}

The quantitative results depend on the asymmetry of the trap,
$a_\parallel/a_\perp = a_\parallel/(\sqrt{2}\ell)$.
The phase diagram for the
system $a_\parallel/\ell = 0.4$ is shown in
Fig.~\ref{fig:phasediagram}, as a function of $g/C_{\rm d}$ and of
$\mu/\hbar\Omega$. The limit $\mu\ll \hbar\Omega$
corresponds to the LLL. The region of
stability of the system is shown by the solid line.
Within the region of stability, we find a series of vortex lattice
groundstates which include the vortex lattices of the
symmetries found in the LLL calculation\cite{nigel05}: triangular and
square vortex lattices, stripe crystal and bubble crystal phases. (We
refer the reader to Ref.~\onlinecite{nigel05} for illustrations of the
particle distributions in these phases.)
The triangular vortex lattice is the lowest energy state for all
values of $g/C_{\rm d}$ to the right of the dashed-dotted line.  This
includes, in particular, all positive values of $g$.  On the left of
the dashed-dotted line a square vortex lattice has energy lower than the
triangular one.  As $g/C_{\rm d}$ is further decreased the ground
states are stripe crystal phases (in the region between the
dotted and the dashed line) which can also be viewed as
rectangular lattices of the vortices\cite{nigel05}.  We solve for the
ground states in this regime by varying the aspect ratio of the
simulation cell and optimising the period of the stripes.  As
$g/C_{\rm d}$ is decreased the aspect ratio of the rectangular lattice
varies, causing the separation between the stripes to become of larger
period.  For values of $g/C_{\rm d}$ to the left of the dashed line we
find that a bubble crystal phase has the lowest energy.  This is a
triangular lattice of bubbles (clusters of particles), with the
vortices arranged between the bubbles. The simplest bubble crystal
phase has four double vortices per bubble which are arranged on the
sites of the honeycomb lattice that is dual to the triangular lattice
of bubbles\cite{nigel05}.

Our results, presented in Fig.~\ref{fig:phasediagram}, show that the
values of $g/C_{\rm d}$ at which there are transitions between the
vortex lattice phases are only weakly dependent on the chemical
potential, $\mu$. Thus the transitions are closely given by the LLL
theory; consistent with this the particle densities we find (not
shown) closely resemble those in the LLL\cite{nigel05}.
We find that
the main effect of going beyond this weak-interaction regime, is the
appearance of the instability to collapse for large values of
the chemical potential.
The value of $\mu/\hbar\Omega$ at which this instability sets in is a
strong function of $g/C_d$. We associate this strong dependence to the
varying particle density in these phases. The bubble crystal phase has
a very inhomogeneous density, so, for a given average density, the
maximum particle density is much larger in the bubble crystal phase
than in the triangular vortex lattice.
Indeed, in the LLL limit,
$\mu \ll \hbar\Omega$, the particle density at the center of the bubble diverges at
the point of instability, which is $g/C_{\rm d} \to -8.49$ for 
LLL calculation\cite{nigel05} applied to $a_\perp/\ell = 0.4$.
\cite{footnote4} 

Our results show that 
for large values of the chemical potential
only the triangular vortex lattice is stable.
In this limit, $\mu \gg \hbar\Omega$, the
vortex cores are small compared to the vortex spacing,
and the interactions between the vortices arise from the kinetic
energy of the flow around the vortices. This leads to a logarithmic
repulsion between the vortices,\cite{donnelly} and thus a triangular
lattice groundstate.

An estimate for the region of stability of the triangular vortex
lattice in the strongly interacting regime, $\mu/\hbar\Omega\gg 1$,
can be obtained by
considering the condition for stability of the
gas in the regions between the vortices. 
Denoting the spacing of the triangular vortex lattice by $a_{\rm v} =
[(4\pi)^{1/2}/{3}^{1/4}]\ell$, we look for an instability to modes
with wavevectors larger than $q = \alpha (2\pi/a_{\rm v})$ of a
homogeneous Bose gas moving with local velocity $\beta \hbar/(M a_{\rm
v})$ ($\alpha$ and $\beta$ are numerical factors of order one). An
analysis of the Bogoliubov spectrum (discussed in more detail below)
shows that an instability occurs for
\begin{equation}
\frac{\mu}{\hbar\Omega} > \frac{A}{B\frac{a_\parallel}{\ell} \frac{C_{\rm
d}}{g + 4\pi C_{\rm d}} -1},
\label{eq:bound}
\end{equation}
where $A = \sqrt{3}/(2\pi)[(2\pi)^2\alpha^2-2\beta^2]$ and
$B=(2\pi)^2{3}^{1/4}\alpha/\sqrt{2}$.  The functional form (\ref{eq:bound}) provides a
good description of the boundary of stability of the triangular
lattice in Fig.~\ref{fig:phasediagram} at large $\mu/\hbar\Omega$ for
reasonable choices of $\alpha$ and $\beta$. This theory shows that,
for large $\mu/\hbar\Omega\gg 1$, the boundary of stability of the
triangular lattice tends to a fixed value $g/C_{\rm d} = -4\pi + B
\frac{a_\parallel}{\ell}$ that increases with $a_\parallel/\ell$. (To
determine the precise value of this boundary would
require a full solution of the Bogoliubov
modes for the triangular vortex lattice in this regime.)

We have investigated the effect of the trap asymmetry by repeating our
numerical calculations for $a_\|/\ell=0.8$ (a more symmetrical trap).
The results for the region of stability, and the phase boundary between
triangular and square vortex lattices are qualitatively similar to
the case of Fig.~\ref{fig:phasediagram}.
We find now that the transitions between
the triangular and the square vortex lattices occur for larger values of
$g/C_{\rm d} \simeq -2.6$ as compared to the case $a_\|/\ell=0.4$.
This is consistent with results in the LLL, and with the qualitative
behaviour of (\ref{eq:bound}).  Furthermore, the new vortex lattices
are stable up to larger values of $\mu/\hbar\Omega$
(up to  $\mu/\hbar\Omega\simeq 12$ for the square lattice) for this more
symmetrical (less oblate) trap.

In order to observe transitions in the vortex
lattice groundstates induced by dipolar interactions, it is necessary
both to tune $g/C_{\rm d}$ to a sufficiently negative value, using for
example a Feshbach resonance, and also to ensure that the particle
density is sufficiently dilute (or interactions sufficiently weak)
that $\mu/\hbar\Omega$ is sufficiently small for the new vortex
lattice states to be stable. We see in Fig.~\ref{fig:phasediagram}
that this requires
$\mu/\hbar\Omega\lesssim 8$ and $g/C_{\rm d}\lesssim -4.5$ for the
square lattice to be observed.
(Stability of the stripe
crystal and bubble crystal requires more stringent
conditions.)
Our numerical calculation gives directly the number of atoms per vortex
that correspond to the above conditions.
For the parameters relevant for
$^{52}$Cr\cite{griesmaier05}, and choosing $\Omega = 2\pi\times
100\mbox{Hz}$, the above conditions
require that the number of particles per vortex be
$\nu \equiv n_{\rm 2d}/n_{\rm v} \lesssim 1300$.
Similarly, for
$a_\parallel/\ell = 0.8$, the conditions for stability of
the square vortex lattice
($\mu/\hbar\Omega \lesssim 12, g/C_{\rm d}\lesssim -2.5$) lead to
$\nu \lesssim 3000$.
While it is challenging to achieve a vortex lattice in a sufficiently
dilute atomic gas, we note that these conditions
are in well excess
of what has been achieved in rapidly rotating Rubidium
condensates.\cite{schweikhard04,bretin04}

Previous calculations of the effects of dipolar interactions on vortex
lattices performed in a {\it trap} geometry\cite{yi05} found direct
transitions from triangular vortex lattice to collapse. It is
not straightforward to make direct connections between the present results for
the phase diagram of a {\it homogeneous} vortex lattice, with those
studies on {\it inhomogeneous} trapped systems. However, we note that
in the studies of Ref.~\onlinecite{yi05} the number of vortices is
relatively small, so the confinement induces a sizeable variation in
the particle density even on the scale of a vortex spacing. It is
possible that, owing to the higher particle density at the centre of
the trap, a trapped system with a small number of vortices is less
stable to collapse.  Our results, Fig.~\ref{fig:phasediagram}, apply
to a system with a large number of vortices such that density
variation over the scale of the vortex lattice constant is small.

\section{Non-rotating Dipolar Bose-Gas}

We now turn to investigate the non-rotating gas.
It is natural to ask whether the stripe and bubble states
that appear for a rotating condensate are manifestations
of a ``supersolid'', 
in which density-wave order appears spontaneously,
in the non-rotating groundstate.
In the latter case one would expect its response to
rotation to involve the vortices being placed in regions of low
particle density. Depending on the symmetry of the supersolid state
this could lead to particle distributions
similar to those of the stripe
crystal and bubble crystal states.

A further motivation in this direction if offered by
the development of a roton minimum in the excitation
spectrum of a homogeneous condensate with dipolar interactions
\cite{odell04,santos03} which can touch zero energy at a finite
wavevector. This indicates an instability of the homogeneous Bose gas and
is suggestive of a possible phase transition to a density-wave
state ({\it e.g.} a condensate of Bogoliubov modes with non-zero
wavevector).

The Bogoliubov modes of a quasi-2D homogeneous condensate with dipolar
interactions were discussed in Ref.~\onlinecite{fischer05}. For a gas
with density $n_{\rm 2d}$ the chemical potential is $\mu = n_{\rm 2d}
\tilde{V}_{q=0}$, and the excitation spectrum is $E(q) =
\sqrt{\epsilon_q \left(\epsilon_q + n_{\rm 2d} \tilde{V}_q\right)}$,
where $\epsilon_q = {\hbar^2 q^2}/({2M})$, and
\begin{equation}
\tilde{V}_q = \frac{g+ 4\pi C_{\rm d}}{\sqrt{2\pi}a_\parallel} - 2\pi {C_{\rm d}} \, q
e^{q^2a_\parallel^2/2}\mbox{erfc}(q a_\parallel/\sqrt{2})
\label{eq:ft}
\end{equation}
is the Fourier transform of the effective 2D interaction
% where we have defined ${\tilde g} \equiv g + 4\pi C_{\rm d}$.
\cite{footnote5}.

If, as above, one considers $g$
to be tuned to negative values, the Bogoliubov mode develops a roton
minimum that becomes unstable even for the quasi-2D regime, $\mu \ll
\hbar\omega_\parallel$, for $g/C_{\rm d}= -4\pi + \sqrt{2\pi^3}C_{\rm
d} n_{\rm 2d}/(a_\parallel \hbar\omega_\parallel)$.  (The case $g<0$
was not considered in Ref.~\onlinecite{fischer05}, and consequently
the instability discussed there, for $g\geq 0$, occurs outside of the
regime of validity of the quasi-2D approximation, in the crossover to
the 3D regime $\mu \simeq \hbar\omega_\parallel$.)

We have looked numerically for
supersolid states in the
quasi-2D Bose gas. We set $\Omega=0$ in the code used for the
numerical calculations described above, chose parameters close to the
instability, and looked for stable (or metastable) states of the
system. For all parameter values that we have tested (which are in the
quasi-2D regime), we find that the only stable state is the homogeneous
Bose gas. Configurations with ``supersolid'' density
wave order (showing a triangular lattice of peaks in the density)
are unstable in
all cases that we have studied.
The collapse mechanism involves a process in which all particles
accumulate into one of the peaks, generating a region of the
sample in which the density diverges.  We have tested this conclusion
by independent numerical calculations (based on a discretisation in
momentum space, and minimisation by the conjugate gradient method) and
have found consistent results.
Thus we conclude that, within mean-field theory, there are no stable supersolid phases for an
infinite quasi-2D Bose gas with interaction (\ref{eq:potential}).

\section{Conclusions}

We have determined the phase diagram for a rapidly rotating atomic
Bose gas with dipolar interactions, beyond the weak interaction
regime.  Our results establish that dipole interactions can drive
transitions from the triangular vortex lattice into vortex lattices of
various different symmetries. These provide estimates of the
experimental parameters that are required in order to observe vortex
lattice groundstates of types other than triangular.

We also investigated the fate of a {\it non}-rotating quasi-2D dipolar
Bose gas at the point at which the homogeneous Bose gas becomes
unstable to Bogoliubov modes with a finite wavelength. Our calculations
indicate that, within the mean-field approach used, there do not exist stable
supersolid states, but that the homogeneous Bose gas becomes unstable to collapse.

The stripe and bubble phases discussed above for a rotating gas are
therefore not manifestations of stable supersolid phases of the
non-rotating gas. Rather, one should view the rotation as acting to
{\it stabilise} the density-wave order that the dipolar interactions
tend to induce in the dipolar Bose gas. The rotation introduces a
lengthscale, the magnetic length $\ell$, which, in the weakly
interacting regime $\mu/\hbar\Omega\ll 1$, sets the minimum wavelength
over which the condensate wavefunction can vary. This minimum
lengthscale prevents collapse of the system provided
interactions are sufficiently weak.

\vskip0.5cm

{\it Note added:} While preparing this work for publication we learned that
G. Shlyapnikov and P. Pedri have reached the same conclusion: that putative supersolid
states in the quasi-2D Bose gas with the interaction (\ref{eq:potential}) are unstable.\cite{shlyapnikov06}

\begin{acknowledgments}
This work was partially supported by EPSRC Grant No. GR/S61263/01 and
by the ICAM Senior Fellowship programme (NRC).  
\end{acknowledgments}


\begin{thebibliography}{10}
\bibitem{footnote1}
Using a
Thomas-Fermi approximation for a harmonically trapped gas with a large
number of vortices, it is simple to show that
$\frac{\omega_\perp^2}{\Omega^2} = 1+
\frac{2\mu}{\hbar\Omega}\frac{1}{N_{\rm v}}$. Thus
$\omega_\perp/\Omega \to 1$ at $N_{\rm v}\to \infty$.
Thus $\omega_\perp/\Omega \to 1$ at $N_{\rm v}\to \infty$.
\bibitem{footnote2}
We define the chemical potential to be
measured relative to the zero-point energy in the rotating frame,
$\hbar\Omega + (1/2) \hbar\omega_\parallel$.
\bibitem{footnote3}
Since the different vortex phases have different unit cells
they cannot all be simulated with the same value of the lattice
discretisation $\Delta x$.  In order to minimize the effects of
discretisation, we compute each for a range of lattices with a range
of $\Delta x$, and interpolate these results so that the energy
comparison is made at the same value of $\Delta x$.
\bibitem{footnote4}
In Fig.\ref{fig:phasediagram} as $\mu\to 0$ we find that the
system is stable for $g/C_{\rm d} > -6.8$. The discrepancy from the
LLL result, $g/C_{\rm d} = -8.49$ is due to our discretization of
the condensate wavefunction in the present work (for finer grids we
find improved agreement with the LLL result), so can be viewed as a
measure of the numerical errors of the present calculation.
\bibitem{footnote5}
Note
that by excluding the point $r=0$ our definition of the dipolar
interaction (\ref{eq:vd}) differs from that of
Ref.~\onlinecite{fischer05} by a contact term (a constant in momentum
space).
\end{thebibliography}
\end{document}